BMC
Research Notes

# *In silico* prediction of protein-protein interactions in human macrophages

Oussema Souiai[1,2*], Fatma Guerfali[1], Slimane Ben Miled[1,3], Christine Brun[2] and Alia Benkahla[1]

## Abstract

**Background:** Protein-protein interaction (PPI) network analyses are highly valuable in deciphering and understanding the intricate organisation of cellular functions. Nevertheless, the majority of available protein-protein interaction networks are context-less, *i.e.* without any reference to the spatial, temporal or physiological conditions in which the interactions may occur. In this work, we are proposing a protocol to infer the most likely protein-protein interaction (PPI) network in human macrophages.

**Results:** We integrated the PPI dataset from the Agile Protein Interaction DataAnalyzer (APID) with different meta-data to infer a contextualized macrophage-specific interactome using a combination of statistical methods. The obtained interactome is enriched in experimentally verified interactions and in proteins involved in macrophage-related biological processes (*i.e.* immune response activation, regulation of apoptosis). As a case study, we used the contextualized interactome to highlight the cellular processes induced upon *Mycobacterium tuberculosis* infection.

**Conclusion:** Our work confirms that contextualizing interactomes improves the biological significance of bioinformatic analyses. More specifically, studying such inferred network rather than focusing at the gene expression level only, is informative on the processes involved in the host response. Indeed, important immune features such as apoptosis are solely highlighted when the spotlight is on the protein interaction level.

**Keywords:** Protein interaction network, Contextualisation, Macrophage, Inference

## Background

Nowadays, infectious respiratory diseases such as tuberculosis (TB) are no longer a major concern for third world countries only. According to the WHO, one third of the worldwide population is infected with Mycobacterium tuberculosis (MTB) in a latent (Latent form Tuberculosis; LTB) and about ten million cases of Active Tuberculosis (ATB) occur annually [1]. The HIV-TB co-infection also plays a major role in the increase of active tuberculosis cases around the world [1]. Although TB is curable by an adequate antibiotic treatment, patient compliance is often problematic and many clinical cases show multi-drug resistance [2]. These cumulated observations underscore the importance of continued investigation into the mechanisms used by the infectious agent, *Mycobacterium tuberculosis*, to persist and overturn inside the host cell. The TB infection mostly occurs by aerosols and MTB infects alveolar macrophages, which then provide an environment for replication and persistence of bacilli. To do so, the bacterium uses several host cellular pathways such as the PI(3)kinase network around PKB/AKT1 [3] to subvert the immune response and to persist into the macrophage. In response, the host activates the same pathway to trigger the elimination of the pathogen [4]. The intricacy of these mechanisms on one hand, and the potential utility of protein-protein interaction (PPI) network analyses to understand the various cellular mechanisms on the other hand, led us to hypothesise that identifying the PPI network in infected macrophages, would provide new insights concerning the infection and the persistence of the pathogen within its host cell. Indeed, PPI are key elements in the organisation of cellular functions [5]. In the post-genomic era, most of these interactions have been identified by either of two high-throughput methods: the yeast two-hybrid (Y2H) system [6] and affinity purification followed by

* Correspondence: souiai@gmail.com
[1]LIVGM + Laboratory of Medical Parasitology, Biotechnology and Biomolecules, Institut Pasteur de Tunis, Avenue Jugurtha, Tunis, Tunisia
[2]TAGC, Inserm UMR_S 1090, Aix-Marseille Université, Marseille, France
Full list of author information is available at the end of the article mass spectrometry (AP-MS) [7]. Numerous methods aiming at inferring interactions have also been proposed, based on sequence signatures and similarities, domain profiling or bayesian predictions [8-11]. Overall, the assembly of all these PPI added to those identified by small-scale experiments, form large networks called 'interactomes' [12]. Bioinformatic analyses of these networks have led to numerous functional insights such as function prediction for uncharacterised proteins [13-18], evolution of the function of the duplicated genes [19-21] and the organisation of the signalling pathways [22,23].

However, it is important to note that these interactomes are devoid of spatio-temporal information. Indeed, interactions identified by the Y2H techniques are biophysically possible but physiologically context-less. They therefore remain hypothetical until their characterisation in particular conditions *in vivo* [24]. In this context, the reconstruction of contextualised macrophage interactome is a crucial methodological step towards a comprehensive study of MTB infection. To support and strengthen the potential occurrence of the interactions discovered using high-throughput and bioinformatic inference methods in particular physiological contexts, additional functional features such as co-expression correlations, genetic interactions, and functional protein annotations have been routinely used as secondary meta-data to contextualize interactomes [25-27] particularly in a bayesian framework [28].

In this work, we propose a contextualised macrophage PPI network resulting from the combination of PPIs with functional annotations and expression data. To achieve this, we used as an initial step, statistical and functional criteria to select a Confidence Subset (CS) of interactions containing those likely occurring *in vivo* in the human macrophage. After showing the reliability of the CS, we used it as a cornerstone to infer the most likely macrophage interactome. The summary of the complete pipeline is illustrated in Figure 1.

We then verified the specificity of the contextualized macrophage interactome composed of 30,182 interactions by showing that it is enriched in proteins related to the immune response, expressed in macrophages according to the Human Protein Atlas [29] and HPRD [30] and belonging to the host regulatory network during MTB infection [31] as well as in interactions reported to occur in macrophages according to InnateDB [32]. As a last step, aiming at pointing towards the modifications of the macrophage interactome induced by MTB exposure, we used the contextualized interactome to highlight the cellular processes at work upon MTB infection. Interestingly, we showed that considering protein interactions rather than differentially expressed genes provides complementary functional information.

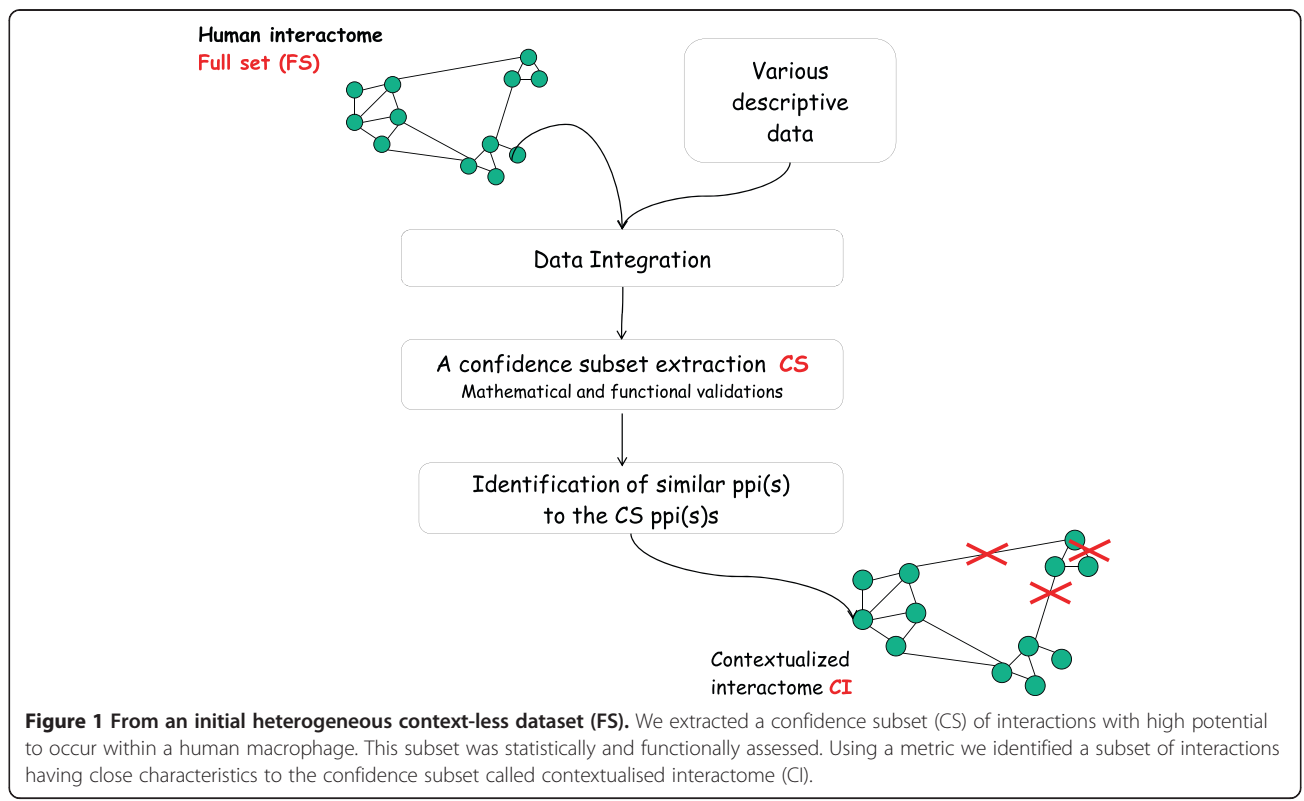

**Figure 1 From an initial heterogeneous context-less dataset (FS).** We extracted a confidence subset (CS) of interactions with high potential to occur within a human macrophage. This subset was statistically and functionally assessed. Using a metric we identified a subset of interactions having close characteristics to the confidence subset called contextualised interactome (CI).

## Results

### Contextualizing the interactome

#### Integrating data to constitute a full dataset

We extracted the human interaction dataset from the APID database [33]. Additional information was integrated to describe each interaction. The following qualitative and quantitative descriptors were used: independent methodological proofs and reports of the interaction, gene co-expression in macrophage, functional co-annotation and sub-cellular co-localisation of the interaction partners (see Material & Methods for the detailed processing of the descriptors). For the sake of clarity, the full dataset composed of the values taken by the descriptors of each interaction was named the Full Set (FS).

### Defining a confidence subset (CS) of macrophage interactions

From the FS composed of 38,832 interactions involving 9,813 proteins, we extracted a Confidence Subset (CS) composed of interactions that likely occur in macrophages, using functional and statistical parameters. For this, we used principal component analysis (PCA) that allows assembling parameters showing similar behaviours (see Material & Methods for details). According to the correlations obtained, the number of reports and evidences are correlated as well as the number of common Gene Ontology terms describing the cellular components and biological processes in which protein pairs are involved (Figure 2). These statistical observations are used to discriminate the CS interactions.

Considering that gene co-expression is routinely used as a parameter in contextualization attempts [25,26], we included only interactions between the products of genes co-expressed in macrophages in the CS. Ultimately, considering that proteins composed of known interacting domains have higher confidence in the PPI network, we selected only interactions between partners sharing interacting domains according to PFAM annotations.

Overall, each interaction belonging to the CS obeys the following criteria (Figure 3, Material & Methods for details):

1) the genes encoding the interacting proteins must be co-expressed in normal, uninfected macrophages;
2) the protein partners must share interacting domains according to PFAM;
3) protein partners must share functional Gene Ontology annotations;
4) the interaction must have been identified several times by independent experiments.

In this way, a CS composed of 530 interactions involving 594 proteins was obtained. The analysis of the Gene Ontology terms annotating those proteins showed that the CS is enriched in terms related to immune system development (Fold Change, FC = 3, p-value = $1.06 \times 10^{-5}$), to regulation of apoptosis processes (FC = 3.29, p-value = $2.2 \times 10^{-26}$), to regulation of cell death (FC = 3.6, p-value = $8.1 \times 10^{-29}$) and to regulation of I-kappaB kinase/NF-kappaB cascade (FC = 5.1, p-value = $5.7 \times 10^{-11}$) (Additional file 1: Table S1).

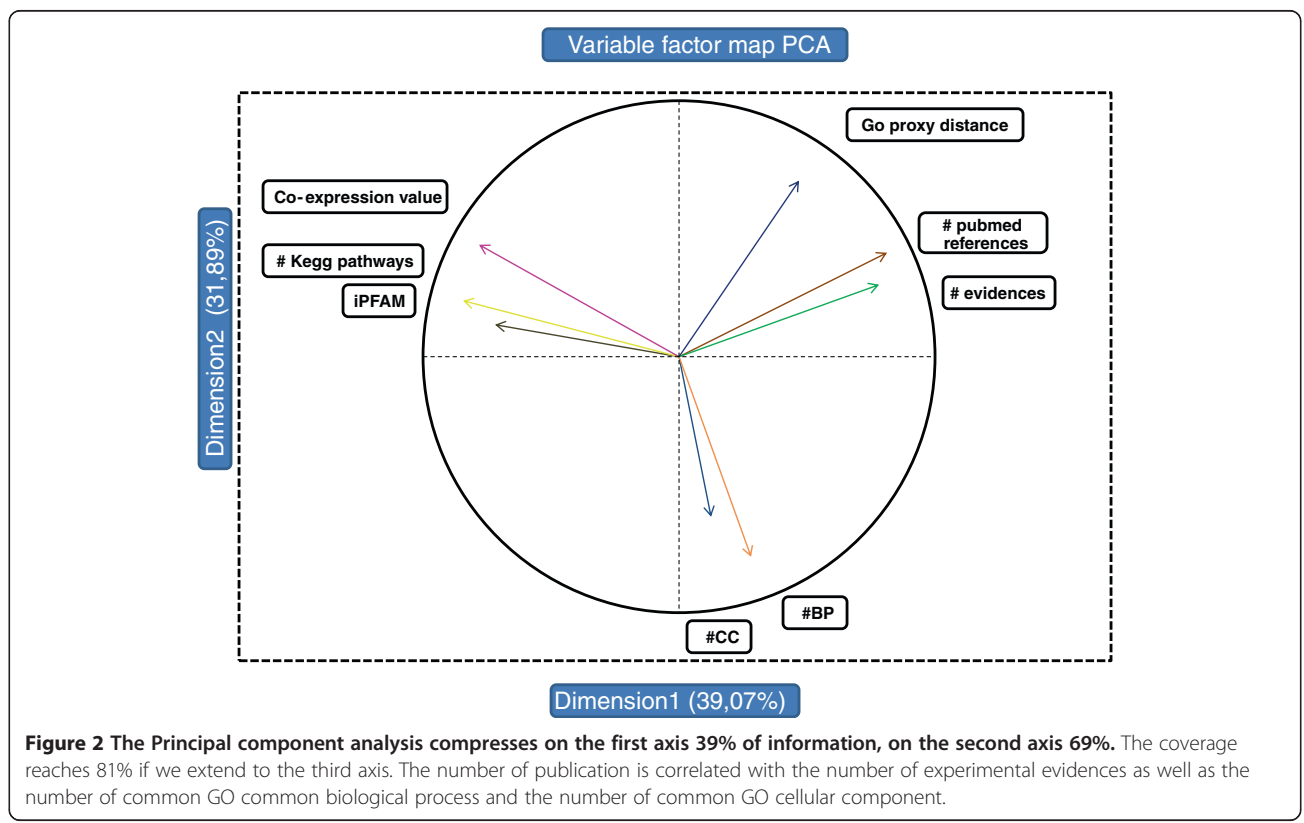

**Figure 2 The Principal component analysis compresses on the first axis 39% of information, on the second axis 69%.** The coverage reaches 81% if we extend to the third axis. The number of publication is correlated with the number of experimental evidences as well as the number of common GO common biological process and the number of common GO cellular component.

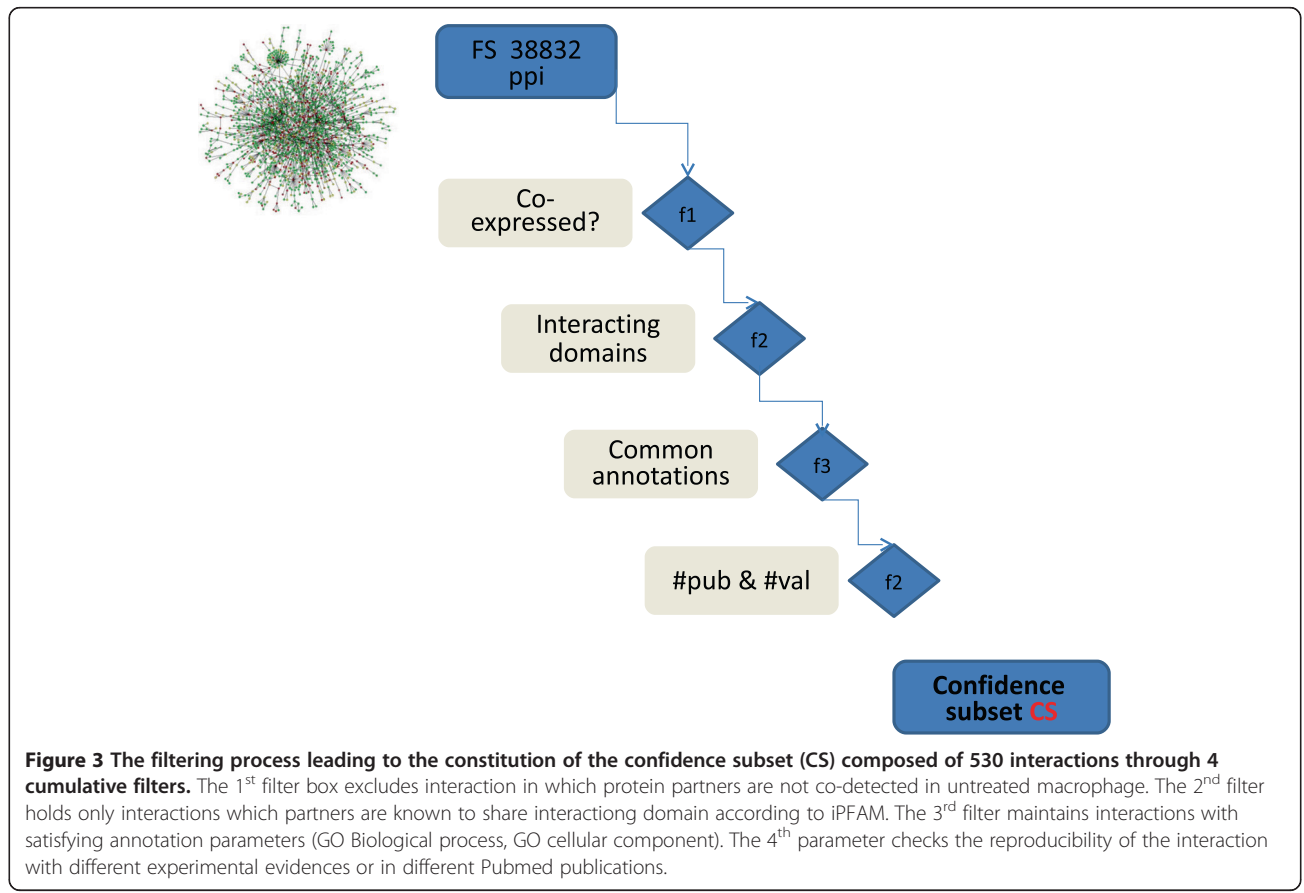

**Figure 3 The filtering process leading to the constitution of the confidence subset (CS) composed of 530 interactions through 4 cumulative filters.** The 1st filter box excludes interaction in which protein partners are not co-detected in untreated macrophage. The 2nd filter holds only interactions which partners are known to share interacting domain according to iPFAM. The 3rd filter maintains interactions with satisfying annotation parameters (GO Biological process, GO cellular component). The 4th parameter checks the reproducibility of the interaction with different experimental evidences or in different Pubmed publications.

Aiming to further gain confidence in the CS, we compared our empirical filtering process to clusters obtained upon applying an unsupervised clustering method to the FS. Interestingly, the Self Organizing Map (SOM) [34] analysis showed that 64% of the interactions contained in the CS are grouped in a single cluster, the remaining interactions being located in 5 out of 16 clusters (Figure 4). This shows that the interactions grouped into the CS according to the criterion empirically chosen (described above) are in agreement with clusters obtained mathematically using an unsupervised algorithm.

In conclusion, the functional enrichment of relevant groups of genes and the satisfactory comparison to unsupervised clustering reinforce the hypothesis that the interactions composing the CS likely occurring in the macrophage.

### Delineating the macrophage protein interaction network

To identify the most likely macrophage PPI network, the interactions most resembling those of the CS were selected by computing a similarity distance. To this end, the CS interactions barycentre was first identified and compared to the descriptor values of the FS interactions. In this case, the barycentre is computed as the centre of mass of all the CS interactions. In other words, considering that CS interactions represents a cloud of points in a multidimensional space with an axis for each of the variables (descriptors), the barycentre of these interactions is defined by the mean of each variable. The barycentre is identified for CS elements as a centroid point whose coordinates represent a vector as follows: $\left[ \frac{1}{n} \sum_{i=1}^{n} CSd_1, \frac{1}{n} \sum_{i=1}^{n} CSd_2, \ldots, \frac{1}{n} \sum_{i=1}^{n} CDd_8 \right]$, where $CSd_i$ represents the confidence subset descriptor index.

Second, we computed and compared the distributions of the Euclidean distance values between the barycentre and the CS interactions on one hand, and the FS interactions on the other hand (Figure 5). We then considered as possible in a macrophage, all the FS interactions showing a distance value to the barycentre less than 4.2, *i.e.* the value corresponding to 95% of the surface of the CS distribution. In other words, this cut-off was used to select a CS-like behaviour among the FS interactions.

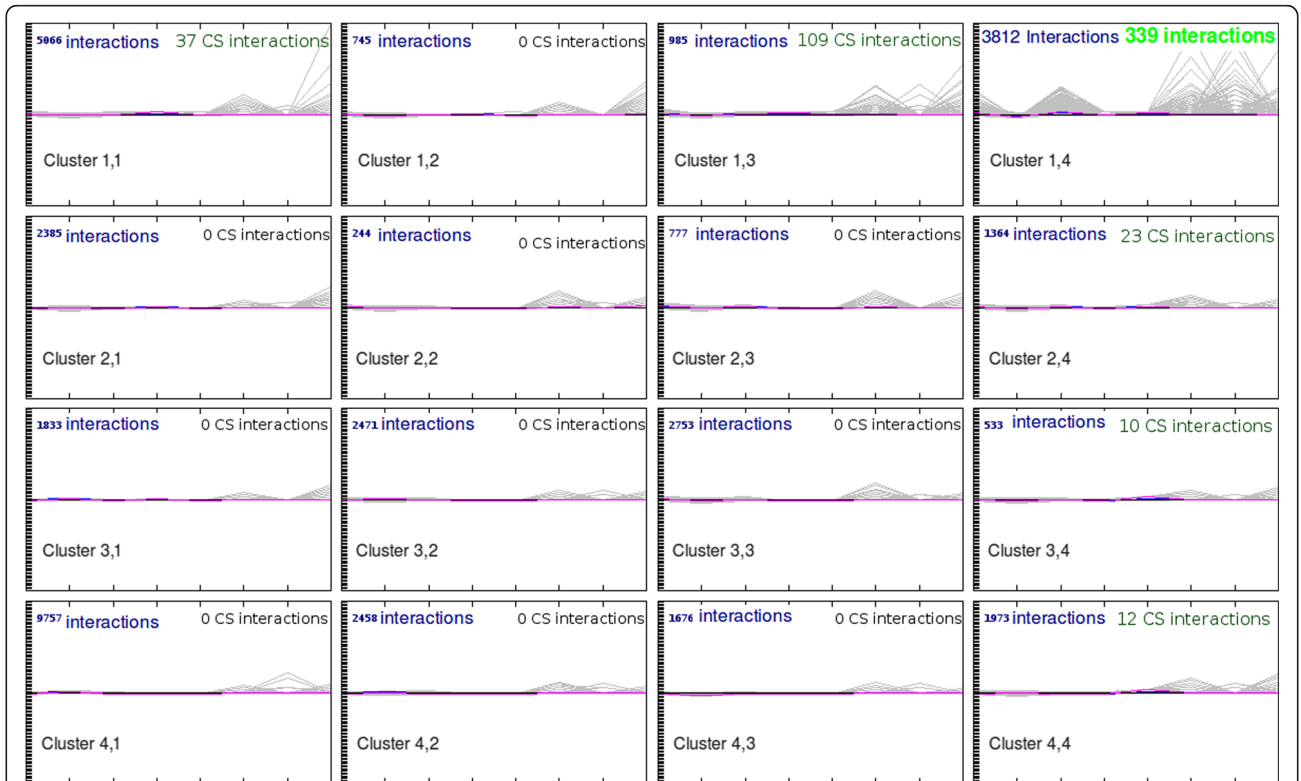

**Figure 4 Self Organizing Map:** The grid (4 4) resulting of the SOM method, applied to the inferred dataset formed by 30182 interactions, shows that the confidence subset is mainly (85% of the CS elements) distributed on 2 neighbor clusters: the node (1, 4) contains 339 elements, the node (1, 3) contains 109 interactions.

The resulting Contextualized Interactome (= CI) is composed of 30,182 interactions involving 8,633 proteins, corresponding to 75% of the initial FS. This ratio can be taken to mean that nearly 75% of the interactions composing an interactome are possible in a given tissue [35].

**Validating the macrophage protein interaction network**

In order to increase our confidence in the contextualisation process, we verified the functional enrichment of the CI compared to the FS. We found that interactions involved in regulation of apoptosis (FC = 1.05, p-value = $1.05 \times 10^{-7}$) and cellular death mechanisms (FC = 1.044 p-value = $2.2 \times 10^{-4}$) (Additional file 2: Table S2) are enriched, underlining the over-representation of pathways involved in the immune response to pathogenic exposures in the CI.

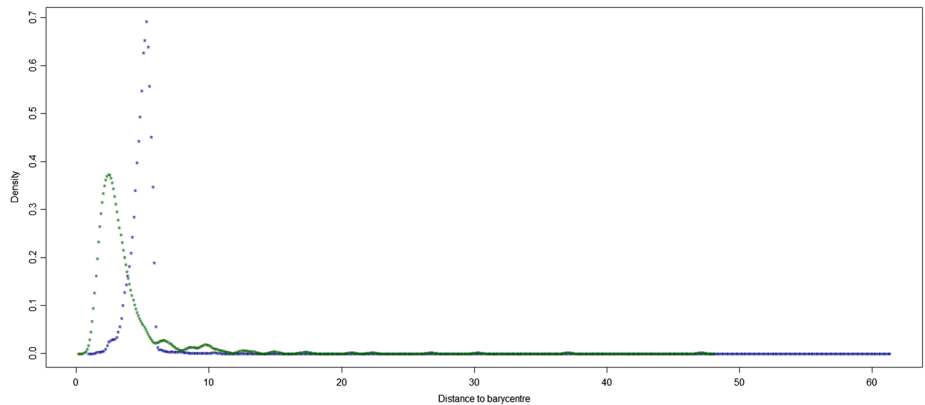

**Figure 5 The density distribution of distances of CS elements to the barycentre (green curve) and the FS elements to the barycentre (blue curve).** The cut-off of CS like elements corresponds to 95% of the surface of the green curve. FS elements having distance lower than 95% the computed threshold were considered as similar to the CS and consequently possible within a macrophage.

Although the CI corresponds to three quarters of the FS, its functional terms are more significantly enriched compared to those of the FS (See Figure 6 and Additional file 3: Table S5). Similarly, the CI observed functional annotations terms are more significantly enriched compared to those obtained from randomized interactomes (Additional file 4: Figure S2).

To further assess, statistically, the resulting CI, we compared the host regulatory network following an exposure to MTB [31] with the CI and with randomly obtained interaction sets. Interestingly, the CI is significantly enriched (p-value < $2.2 \times 10^{-16}$; $t$-test) in interactions reported in the MTB regulatory network compared to random interaction sets. Likewise, the CI is statistically enriched in interactions experimentally identified in macrophages according to InnateDB [32], a database devoted to innate immunity ($t$-test p-value < $2.2 \times 10^{-16}$, see Materiel & Methods for details).

To complement our analysis, we computed the overlap of CI with interactomes contextualised using other sources: the macrophage proteome from Protein Atlas [29] and HPRD [30] (see the Materiel & Methods for details). The CI overlaps satisfactorily with the HPRD macrophage interactome (p-value = 0.022) and more significantly with the Protein Atlas macrophage interactome (p-value = $3.56 \times 10^{-22}$) (see Figure 7).

Altogether, these comparisons summarised in Figure 7 emphasize the "macrophagic" specificity of the contextualized interactome (CI).

## Study case: macrophage cellular processes modulated by a bacillary infection from an interactome point of view

In order to evaluate the pertinence of the contextualized macrophage interactome, we used it in the following study case.

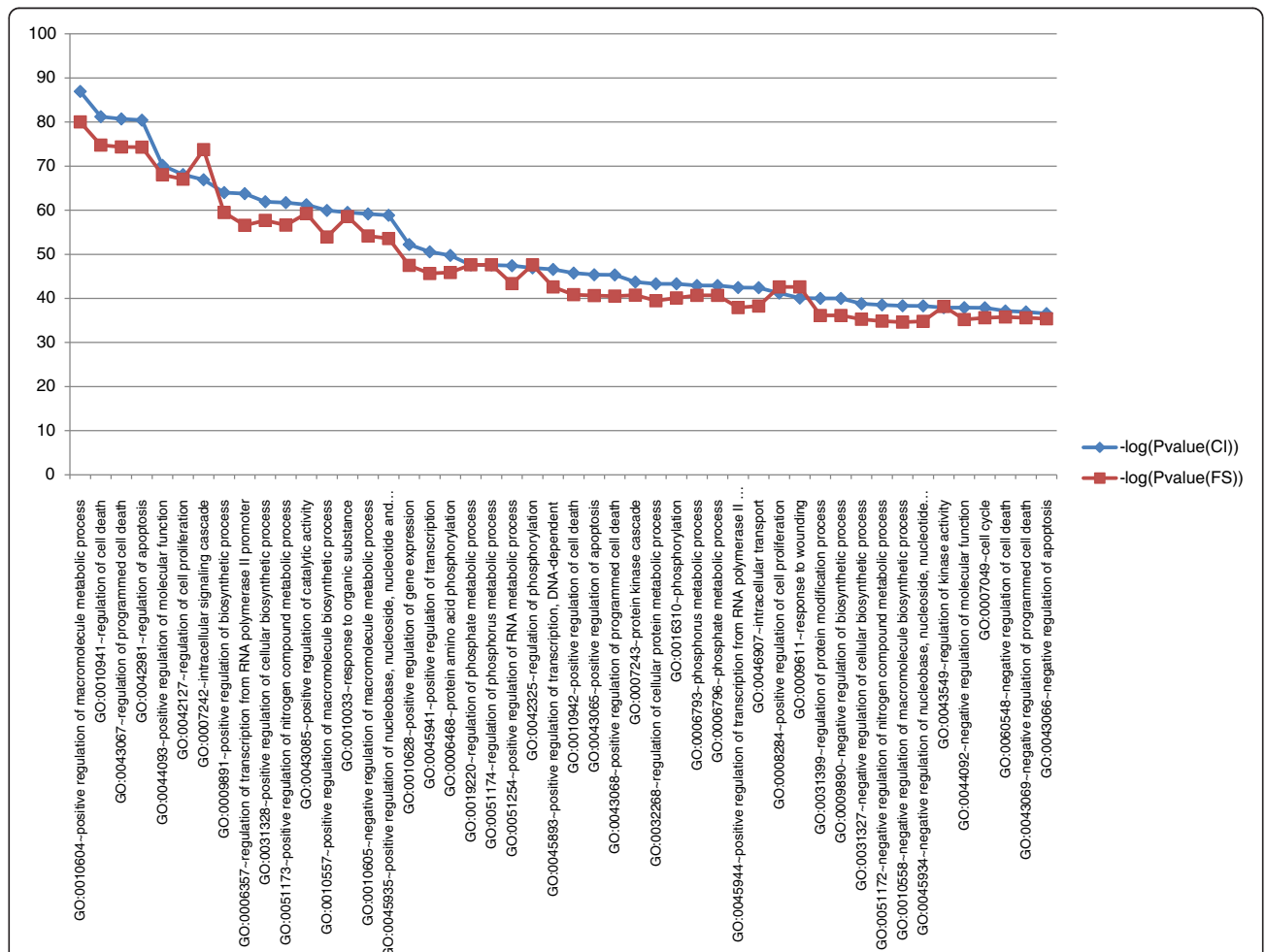

**Figure 6 Top 50 comparison enrichment terms p-values: The CI enrichments p-values (blue line) are more enriched than the FS enrichments p-values (red line).** The difference in enrichment p-values between the two sets is significant according to a $t$-test (df = 499.626, p-value = 0.01587).

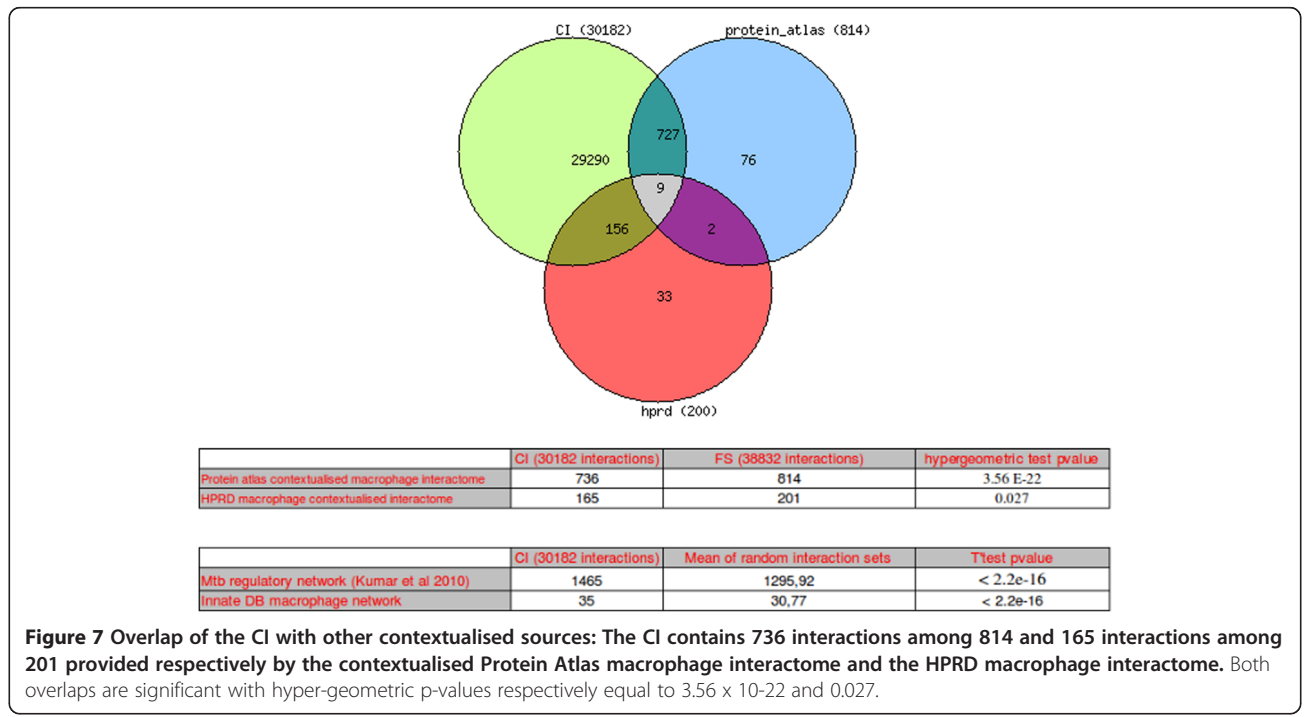

| | CI (30182 interactions) | FS (38832 interactions) | hypergeometric test pvalue |
|---|---|---|---|
| Protein atlas contextualised macrophage interactome | 736 | 814 | 3.56 E-22 |
| HPRD macrophage contextualised interactome | 165 | 201 | 0.027 |

| | CI (30182 interactions) | Mean of random interaction sets | Ttest pvalue |
|---|---|---|---|
| Mtb regulatory network (Kumar et al 2010) | 1465 | 1295,92 | < 2.2e-16 |
| Innate DB macrophage network | 35 | 30,77 | < 2.2e-16 |

**Figure 7 Overlap of the CI with other contextualised sources: The CI contains 736 interactions among 814 and 165 interactions among 201 provided respectively by the contextualised Protein Atlas macrophage interactome and the HPRD macrophage interactome.** Both overlaps are significant with hyper-geometric p-values respectively equal to 3.56 x 10-22 and 0.027.

The expression signatures of macrophages infected with MTB have been characterized in three independent studies [31,36,37]. By combining these data, we obtained two lists of down-regulated and up-regulated genes upon MTB infection. Based on the SAM algorithm (Additional file 5: Figure S1), we have ultimately cumulated 3,724 under-expressed genes and 1,651 over-expressed genes from the three transcriptomic experiments. We focused on these genes, knowing that MTB infection regulates the activity of particular host genes and cellular processes to its own benefits. To evaluate the insights brought by a PPI level analysis *versus* a classical differential gene expression approach, we extended the list of genes revealed by SAM to their first interactors in the CI, thus defining two sub-networks of 2,966 and 1,435 interactions anchored respectively on the 3,724 under-expressed and the 1,651 genes over-expressed upon infection (note that not all the modulated genes have interactions in the CI). We then compared the functional enrichments of the modulated gene lists and their resulting sub-networks. As shown in Additional file 6: Table S3, whereas the GO terms 'response to oxygen levels', 'cell substrate adhesion', 'cell matrix adhesion', 'positive T cell selection' are the most enriched terms when only under-expressed genes are considered, 'regulation of programmed cell death', 'negative and positive regulation of apoptosis' or 'response to wounding' are found to be over-represented when the interactors are taken into account. Similarly, considering the up-regulated genes and their associated sub-network led to the same finding (see Additional file 7: Table S4).

Therefore, focusing on the interactions involving the products of the regulated genes rather than only on the expression of the genes favours the emergence of functional aspects caused by MTB infection. Among these aspects, the regulation of the apoptosis is known to be highly targeted and controlled by the pathogen during the different phase of infection and persistence in the macrophage, as is nicely discussed by Lee and colleagues [38]. Notably, although these regulatory aspects are crucial for the outcome of infection, they are more significantly and extensively revealed at the systemic scale by focusing on the PPI. These findings highlight the need to consider infection of the host by a pathogen at the level of the functional module, defined as a group of interacting proteins involved in the same pathway or biological process, instead of focusing solely on genes or their products.

Moreover, considering the interactome revealed that the products of the down-regulated genes after infection, are closer to each other in the network than the rest of the CI proteins. This supports the hypothesis that MTB targets proteins participating to the same pathways. Indeed, the shortest path values between the down-regulated genes are significantly lower than the shortest path between the CI proteins (Mean paths for CI and down-regulated genes within the CI are respectively 3.3 and 4.5 (p-values 0.002557; *t*-test)).

Overall, these results suggest that the bacillus acts upon key proteins, which are closely connected within the network to regulate the host response.

## Discussion and conclusions

Interactomes are undoubtedly a remarkable means to investigate infectious diseases. By multiplying data types and sources, we are able to increase the pertinence of the downstream conclusions.

In this study, we proposed a method to contextualise the interactome of a particular cell type by integrating diverse information. In the data integration process, the expression correlation is subject of debate. Even though this parameter has been taken into account to propose contextualised interactomes [25], this hypothesis has to be considered carefully. Indeed high mRNA expression levels do not necessarily imply a correlated protein expression level and moreover, do not imply the interaction between partner proteins [39].

An interaction requires the presence of both interacting proteins for its accomplishment. This condition is necessary but not sufficient. In the competitive cellular environment, the occurrence of a particular interaction rather than another possible interaction depends on physico-chemical factors (temperature, pH, covalent modifications such as phosphorylation) [40]. These observations have to be taken into consideration to improve the contextualisation process. Nevertheless, although integrating tissue and cell type information into interaction network is certainly a desirable goal (see discussion of [41]), few attempts have been reported. Interestingly, only a few types of data were integrated at one time: Bossi and Lehner [25] proposed tissue specific interactomes by integrating gene expression and PPI showing that most 'housekeeping' proteins have important tissue-specific interactions; similarly, Rachlin and colleagues [27] provided networks dedicated to particular biological processes by contextualizing them with Gene Ontology terms. The multiplicity of the integrated data sources was also brought together in a bayesian framework, aiming at proposing functional maps to help the user to build functional hypothesis [28] and in the analysis of a diverse collection of genome-wide data sets (gene expression, protein interactions, growth phenotype data, and transcription factor binding) to decipher the yeast system modular organisation [42]. Our approach relies on the fact that we used multiple sources of data in order to be able to propose a tissue-specific network of high confidence. The use of multiple data descriptors offers a global view and aims at minimizing the biases for interactome contextualisation.

Second, we used a learning approach based on the constitution of a statistically and functionally reliable CS

in order to select the interactions likely to occur in a macrophage. Contextualising networks and defining dense sub-networks and functional modules governing the host response to infection offers a complementary approach to classical analysis for the investigation of infectious diseases. Moreover, considering the modular composition of the host interactomes allows inclusion in the analyses of major actors of the immune response and maintenance of cell fate that would not have been tractable if considering gene or protein data alone. Overall, our work suggests that contextualizing interactomes improves the biological significance of bioinformatics analyses.

## Methods

### Human interactome descriptors

From APID we extracted an interactome dataset composed of 38832 interactions involving 9831 proteins. Features were added to compose a dataset of interactions described by functional and quantitative descriptors:

1. **# methods**: This information is extracted from APID and corresponds to the number of experimental validations describing the interaction according to the molecular interaction controlled vocabulary PSI-MI [43]. Only leaves of the PSI-MI experimental validation tree were selected.
2. **# publications**: extracted from APID. Corresponds to the number of articles indexed in PubMed and reporting the interaction.
3. **iPFAM value**: extracted from APID. Identifies whether the interactors pair contains domains known as interacting according to the Pfam database [44].
4. **GO-proxy**: this program is part of the GOToolBox suite [45]. It computes a similarity index between the interactors on the basis of the GO annotation terms they share. The similarity index corresponds to Czekanowski-Dice formula [13,46].
5. **# of common GO biological process terms**: represents the number of common GO biological processes shared by the interacting proteins. For sake of precision, we only consider terms found at level 3 in the ontology tree.
6. **# of common GO cellular component terms**: corresponds to the number of common GO cellular components shared by interactors.
7. **# of common KEGG pathways**: corresponds to the number of KEGG pathways shared by the interactors.
8. **Co-expression value**: macrophage expression data from Chaussabel and colleagues [36], downloaded from the Gene Expression Omnibus database (GEO) [47]. Each probe set corresponds to a mRNA and was categorized either by Present, Absent or Marginal. The Presence/Absence call of the mRNA was calculated according to the MAS5.0 algorithm [48].

To evaluate the occurrence of the interaction considering the Presence/Absence status of the mRNA, we assumed the following hypotheses:

i. the presence of the mRNA implies the presence of the corresponding protein: the mRNA is detected as present according to the SAM algorithm [49];
ii. for a couple of proteins interacting *in vitro*, if both proteins are considered as present within a targeted cell according to the hypothesis (i), we assume that the interaction is bio-physically possible in that condition.

### Enrichment/depletion analysis parameters

The functional analysis webtool from DAVID (http://david.abcc.ncifcrf.gov/) [50] was used to statistically investigate the terms over-/under-represented in the set of proteins belonging to the CS and the CI. The human genome was used as reference to compare the FS and the CI enrichments (Additional file 3: Table S5).

The set of proteins composing the FS interactions was used as reference to compute the enrichment of the CS (Additional file 1: Table S1) and the enrichment of the CI (Additional file 2: Table S2).

The set of proteins composing the CI interactions was used as reference to compute the enrichment of the sub-networks of down-regulated genes and their first interactors (Additional file 6: Table S3) and the enrichment of the sub-networks of up-regulated genes and their first interactors (Additional file 7: Table S4).

The p-values were calculated using a hyper-geometric law and corrected for multi-testing with the Benjamini and Hochberg correction.

### Confidence subset statistical relevance

The CS relevance was assessed by using two distinct clustering algorithms.

### Self organizing Map (SOM)

We used an unsupervised neural network method, the Self-Organizing Map (SOM) [34] for clustering and visualising the high-dimensional complex inferred data on a single map. We applied a Euclidean SOM to the APID original dataset composed of 38832 interactions, with the following parameters: map size $5 \times 10$, Gaussian as neighbour, linear initialisation and rectangular topology. The subset composed of 530 interactions was distributed on three neighbouring clusters. The first one contains 437 interactions, the second contains 83 and the third 10.

### Principal component analysis (PCA)

The R graphical library Rcmdr was used to import and normalise the FS.

This PCA allowed summarising 81% of the global information.

**Contextualised interactomes**: We compared the CI to other contextualised macrophage interactome from various data sources:

**Protein atlas contextualised interactomes**: We queried Protein Atlas [29] (http://www.proteinatlas.org/), to extract a list of proteins having a strong expression in macrophages (1990). To generate a contextualized interactome, we retained only the interactions of the FS between proteins pairs having a macrophage protein expression.

**HPRD macrophage interactome**: From HPRD database (HPRD_Release9_041310), we selected a subset of proteins localised in the macrophage. We finally obtained 201 interactions between interacting partners both localised in the macrophage based on the tissular expression field of the database.

### Additional files

**Additional file 1: Table S1.** Enrichment analysis of the Confidence subset (CS) using the The FS as reference.

**Additional file 2: Table S2.** Enrichment analysis of the Contextualized interactome (CI) using the The FS as reference.

**Additional file 3: Table S5.** Enrichment analysis of the Confidence subset (CI) and the FS as reference using the genome as reference.

**Additional file 4: Figure S2.** Top 50 comparison enrichment terms p-values between CI and five randomised CI(s): The CI enrichments p-values (black line) are more enriched than the observed randomised CI enrichments p-values (p1, p2, p3, p4 and p5). T-test comparisons were performed between the CI and each randomised set of interactions (p1 to p5). The difference remains significant in each case with t-test p-values varying from 4.166e-05(p2) to 0.03316(p1).

**Additional file 5: Figure S1.** Constitution of down-regulated and up-regulated gene sets. These genes were identified through SAM analysis (Significance analysis of microarray) with respect to median false discovery rate of 1%. Red points correspond to up-regulated genes and green points correspond to down-regulated genes. Top analysis [37]; Medium analysis [31]; Bottom analysis [36]. Ultimately these analyses allowed respectively the constitution of respectively 3724 and 1651 up-regulated and down-regulated gene sets.

**Additional file 6: Table S3.** Enrichment of the sub-networks of down-regulated genes and their first interactors.

**Additional file 7: Table S4.** Enrichment of the sub-networks of up-regulated genes and their first interactors.

**Authors' contributions**
OS compiled data analyzed interactomes and wrote the initial draft under the supervision of CB and AB. FG and SM collaborated respectively to transcriptomic and statistical analyses. CB and AB reviewed the final manuscript. All authors read and approved the final manuscript.

## Acknowledgments
We would like to acknowledge Colin Tinsley for his contributions in English spelling and grammar revisions and Javier De Las Rivas for making APID data available.

## Author details
[1]LIVGM + Laboratory of Medical Parasitology, Biotechnology and Biomolecules, Institut Pasteur de Tunis, Avenue Jugurtha, Tunis, Tunisia. [2]TAGC, Inserm UMR_S 1090, Aix-Marseille Université, Marseille, France. [3]ENIT-LAMSIN BP 37, Tunis, Tunisia.